\documentclass[aps,pra,twocolumn,superscriptaddress]{revtex4-1}

\usepackage{graphicx}
\usepackage{hyperref}
\usepackage{amsmath}
\usepackage{amssymb}
\usepackage{epstopdf}
\usepackage{multirow}
\usepackage{threeparttable}
\usepackage[usenames]{color}

\begin{document}

\title{Mid-Infrared Single-Photon Compressive Spectroscopy}
\author{Ben Sun}
\affiliation{State Key Laboratory of Precision Spectroscopy, East China Normal University, Shanghai 200062, China}

\author{Kun Huang}
\email{khuang@lps.ecnu.edu.cn}
\affiliation{State Key Laboratory of Precision Spectroscopy, East China Normal University, Shanghai 200062, China}
\affiliation{Chongqing Key Laboratory of Precision Optics, Chongqing Institute of East China Normal University, Chongqing 401121, China}
\affiliation{Collaborative Innovation Center of Extreme Optics, Shanxi University, Taiyuan, Shanxi 030006, China}

\author{Huijie Ma}
\affiliation{State Key Laboratory of Precision Spectroscopy, East China Normal University, Shanghai 200062, China}

\author{Jianan Fang}
\affiliation{State Key Laboratory of Precision Spectroscopy, East China Normal University, Shanghai 200062, China}

\author{Tingting Zheng}
\affiliation{State Key Laboratory of Precision Spectroscopy, East China Normal University, Shanghai 200062, China}

\author{Ruiyang Qin}
\affiliation{State Key Laboratory of Precision Spectroscopy, East China Normal University, Shanghai 200062, China}

\author{Yongyuan Chu}
\affiliation{Key Laboratory of Specialty Fiber Optics and Optical Access Networks, Shanghai University, Shanghai 200444, China}

\author{Hairun Guo}
\affiliation{Key Laboratory of Specialty Fiber Optics and Optical Access Networks, Shanghai University, Shanghai 200444, China}

\author{Yan Liang}
\affiliation{School of Optical Electrical and Computer Engineering, University of Shanghai for Science and Technology, Shanghai 200093, China}

\author{Heping Zeng}
\email{hpzeng@phy.ecnu.edu.cn}
\affiliation{State Key Laboratory of Precision Spectroscopy, East China Normal University, Shanghai 200062, China}
\affiliation{Chongqing Key Laboratory of Precision Optics, Chongqing Institute of East China Normal University, Chongqing 401121, China}
\affiliation{Shanghai Research Center for Quantum Sciences, Shanghai 201315, China}
\affiliation{Chongqing Institute for Brain and Intelligence, Guangyang Bay Laboratory, Chongqing, 400064, China}

\begin{abstract}
Sensitive mid-infrared (MIR) spectroscopy plays an indispensable role in various photon-starved conditions. However, the detection sensitivity of conventional MIR spectrometers is severely limited by excessive noises of the involved infrared sensors, especially for multi-pixel arrays in parallel spectral acquisition. Here, we devise and implement an ultra-sensitive MIR single-pixel spectrometer, which relies on high-fidelity spectral upconversion and wavelength-encoding compressive measurement. Specifically, a MIR nanophotonic supercontinuum from 3.1 to 3.9 $\mu$m is nonlinearly converted to the near-infrared band via synchronous chirped-pulse pumping, which facilitates both the precise spectral mapping and sensitive upconversion detection. The upconverted signal is then spatially dispersed onto a programmable digital micromirror device, before being registered by a single-element silicon detector. Consequently, the spectral information can be deciphered from the correlation between encoded patterns and recorded measurements, which results in a spectral resolution of 0.5 cm$^{-1}$ under an illumination flux down to 0.01 photons/nm/pulse. Moreover, we demonstrate faithful reconstructions at sub-Nyquist sampling rates by using the compressive sensing algorithm, which leads to a 95\% reduction in data acquisition time. The presented single-pixel computational spectrometer features wavelength multiplexing, high throughput, and efficient sampling, which thus paves a new way for sensitive and fast spectroscopic analysis at the single-photon level.

\end{abstract}

\maketitle

\section{Introduction}
Mid-infrared (MIR) spectral region is highly attractive for molecular absorption spectroscopy due to the accommodation of abundant fingerprinting rotational-vibrational transitions \cite{Vodopyanov2020Book}. Nowadays, MIR spectrometers have been established as a powerful tool for material characterization and identification in various fields, including physics, chemistry, and medical sciences \cite{Griffiths2007Book, Haas2016ARAC, Cheng2015Science}. In particular, high-sensitivity MIR spectroscopy is highly demanded in many applications at photon-scarce conditions, such as remote sensing at a long stand-off distance \cite{Coburn2018Optica}, heritage examination with a limited radiation dose \cite{Daffara2018OLE}, and biological observation under a minimal phototoxicity \cite{Shi2020NM}. Consequently, spectral measurements at high signal-to-noise ratios (SNRs) necessitate the requirement of sensitive MIR detection, which imposes a prominent challenge for existing infrared detectors, especially at room-temperature operation \cite{Razeghi2014RPP}. Indeed, conventional MIR sensors based on narrow-bandgap semiconductors are typically susceptible to intrinsic dark current and ambient thermal perturbation \cite{Wang2019Small}, which may lead to signal distortion or measurement inaccuracy under a low-light-flux illumination. Notably, superconducting nanowire detectors have shown great promise in single-photon infrared spectroscopy \cite{Kong2021NL, Xiao2022ACS}, albeit with system complexity at cryogenic operation and fabrication difficulty in array integration. Recently, emerging platforms based on low-dimensional materials \cite{Keuleyan2011NP, Yuan2021NP, Yu2018NC, Xue2023LSA} hold significant potential in MIR sensing at room temperature, despite with a current sensitivity far away from the single-photon level \cite{Wang2019Small, Chen2020AMT}. Therefore, it is imperative to develop novel techniques for implementing a room-temperature MIR spectrometer with desirable features of wide spectral coverage, high detection sensitivity, and fast refreshing rate.

In the context, the frequency upconversion strategy has attracted increasing attention to circumvent the aforementioned limitations for infrared detectors \cite{Barh2019AOP, Rehain2020NC, Fang2024NC}, wherein the MIR signal is nonlinearly converted to the visible or near-infrared bands for leveraging the high-performance silicon photodetectors. Such upconversion detectors facilitate the demonstrations of highly-sensitive MIR detection \cite{Huang2021PR, Liu2022Optica, Ge2024SA} and imaging \cite{Dam2012NP, Huang2022NC, Wang2023NC} at the single-photon level. Furthermore, the combination of broadband spectral conversion and subsequent wavelength resolving enables the realization of sensitive and fast MIR upconversion spectrometers \cite{Rodrigo2021LPR, Zheng2023LPR, Jonusas2024OE, Zhao2023NC, Cai2022PR}. Generally, the spectral components are resolved based on either a spatially dispersive operation \cite{Rodrigo2021LPR, Zheng2023LPR, Kalashnikov2016NP} or a tunable filtering stage \cite{Cai2022PR}. The former grating-based spectrometer typically requires a multi-pixel detector array, which is costly and restricted with a limited format scale, especially for single-photon detectors \cite{Hadfield2023Optica}. In comparison, the latter monochromator-type spectrometer permits high-resolution spectral analysis via a single-element detector, but usually suffers from a low energy utilization efficiency due to the involved narrow slit \cite{Cai2022PR}. Very recently, another efficient approach for single-pixel upconversion spectrometer is investigated based on the time-stretch configuration \cite{Sun2024LPR, Cai2024SA}, which favors parallel recording of all spectral components in the time domain \cite{Hashimoto2023LSA}. Similar to the spatial counterpart, the spectral intensity at each resolved element becomes extremely weak in the case of low-light-level illumination, which severely limits the SNRs in the single-photon spectral measurement. Alternatively, spectroscopic nonlinear interferometers have been demonstrated to retrieve infrared spectral information in a wavelength-multiplexing fashion by using single-pixel detection and interferometry-based analysis \cite{Lindner2021OE, Tashima2024Optica}. However, a remaining concern lies in the need of mechanical scanning for obtaining the interferograms, which inevitably results in additional system complexity and increased acquisition time \cite{Griffiths2007Book}. Notably, advanced designs have recently been proposed to implement rapid-scan FTIR \cite{Suss2016RSI, Hashimoto2021LPR}, albeit with reduced spectral resolutions and/or narrowed spectral bandwidths. To date, it remains appealing to realize a high-speed broadband MIR spectrograph along with desirable features of polychromatic multiplexing and optical throughput.

Apart from interferometric methods, wavelength multiplexing measurements can be also achieved by resorting to the so-called Hadamard transform spectroscopy \cite{Lim2021OE}, which combines spatial modulation of dispersed light and single-pixel detector. Nowadays, the involved slit masks have commonly been implemented by using digital micromirror devices (DMDs) that offer high pixel resolution, flexible pattern programming, and fast switching speed \cite{Gattinger2021Sensors, Starling2016AO}. The spectral-coding architecture is analogous to single-pixel imaging in the spatial domain \cite{Edgar2019NP}, which favors high sensitivity, fast response, and reduced costs. Similar to the spatial counterpart, the single-pixel spectroscopic paradigm is compatible to the compressive sensing algorithms for performing adaptive and smart sensing at a sub-Nyquist sampling rate \cite{Phillips2017SA, Gamez2016JAAS, Spencer2016NC}, which allows a significant reduction of acquisition time without considerably sacrificing the spectral resolution. However, the single-pixel spectrometers mostly operate in the visible or near-infrared regions \cite{Gattinger2021Sensors, Starling2016AO}, which is usually dictated by the operation wavelength range for conventional spatial light modulators (SLMs). Although a customized DMD with a CaF$_2$ window permits the light modulation at MIR wavelengths \cite{Gattinger2022OE, Ebner2023SR}, the beam steering ability is inevitably degraded by the parasitic diffraction effects especially at longer infrared wavelengths \cite{Edgar2019NP}. Consequently, the spectral resolution for the reported MIR single-pixel spectrometers are typically limited to tens of cm$^{-1}$ \cite{Gattinger2022OE, Ebner2023SR}. So far, MIR single-pixel computational spectroscopy with high sensitivity and high resolution has not yet been realized, which hence urgently calls for the development of techniques to address the challenges on the single-photon detection and high-resolution modulation at MIR wavelengths.

In this work, we propose and implement a MIR single-pixel computational spectrometer based on broadband spectral mapping and sensitive upconversion detection, which offers superior performances with sub-wavenumber resolution and single-photon sensitivity. Here, the upconversion spectrometer nonlinearly converts the MIR radiation into the near-infrared band, where high-performance spatial modulators and photon detectors are leveraged to overcome the long-standing difficulties in high-fidelity infrared manipulation and sensitive sensing at room-temperature. Specifically, a nanophotonic MIR supercontinuum with a broad spectral coverage of 3100-3900 nm is synchronously pumped by a temporally-chirped pump pulse within a nonlinear crystal, which allows a high-efficiency and low-noise spectral transduction while precisely preserving the spectral information in the upconversion replica. Then, the upconversion light is spatially dispersed onto a programmed DMD to perform a high-fidelity spectral encoding. The modulated beams with a series of ordered Hadamard patterns are collected by a single-element silicon photodiode. Finally, the MIR absorption spectrum can be reconstructed based on the knowledge of the projection patterns and the associated measured intensities, which leads to a spectral resolution of 0.5 cm$^{-1}$ at the presence of an incident illumination flux of  0.01 photons/nm/pulse. In combination with a compressive sensing algorithm, a sub-Nyquist sampling strategy is permitted to significantly reduce the data acquisition time by 95\%, which particularly essential for low-light-level spectral analysis in photon-starved scenarios. Notably, the presented MIR compressive single-pixel spectrometer benefits from advantages of both the dispersive and interferometric configurations, thus featuring robust operation, high throughput, and wavelength multiplex. The achieved detection sensitivity here is enhanced by over one order of magnitude better than that for previous MIR upconversion instantiations.
 
\section{Basic principle}
Generally, the proposed MIR single-pixel spectroscopy consists of three core parts: high-fidelity frequency upconversion, high-precision wavelength encoding, and high-accuracy spectrum reconstruction. The frequency conversion is an essential intermediate step for the MIR spectrometer, which is required to translate the broadband infrared radiation in a high-efficiency, low-noise, and high-fidelity manner \cite{Rodrigo2021LPR, Zheng2023LPR}. To this end, a pulsed pumping configuration is used for leveraging the high peak power and narrow pulse duration \cite{Huang2021PR}, which improves the conversion efficiency and suppresses the background noise in comparison to the continuous-wave pump \cite{Dam2012NP, Rodrigo2021LPR}. Moreover, the pump pulse is temporally dispersed to implement the so-called chirped pulse upconversion \cite{Jonusas2024OE, Sun2024LPR}, where the ultrafast MIR signal only overlaps with a small fraction of the pump spectrum in the time domain. The resultant narrow-band pumping ensures a precise wavelength mapping during the nonlinear parametric interaction, which makes it possible to access a high spectral resolution beyond the intrinsic pump bandwidth \cite{Sun2024LPR}.

\begin{figure*}[t!]
\includegraphics[width=0.9\textwidth]{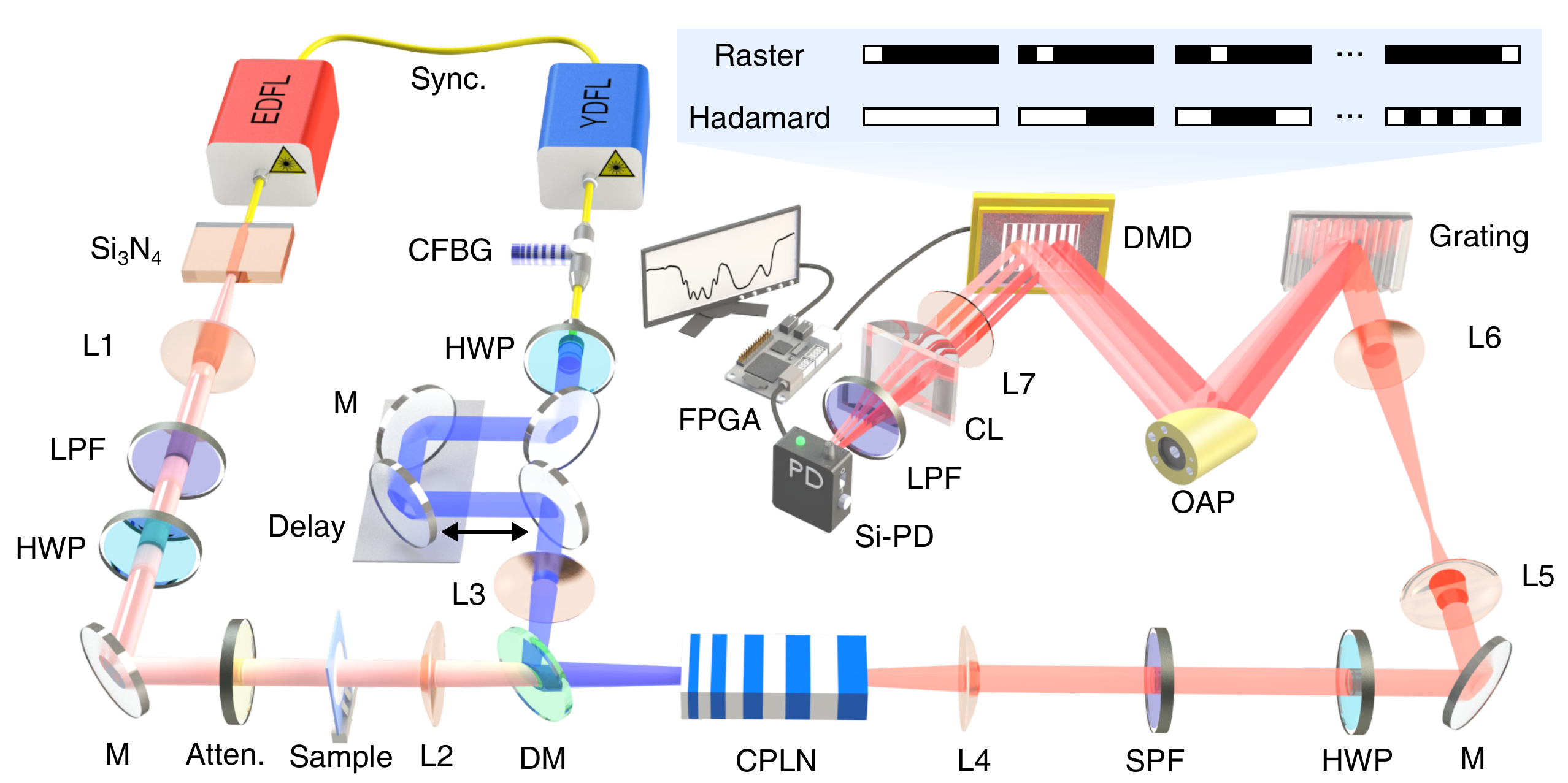}
\caption{Experimental setup of the single-pixel MIR spectral-coding spectroscopy. The involved light sources originate from two passively synchronized Er- and Yb-doped fiber lasers (EDFL and YDFL), which deliver mode-locked pulses at 1550 and 1030 nm, respectively. The femtosecond pulses from the EDFL are launched into a nanophotonic Si$_3$N$_4$ waveguide for preparing a broadband MIR supercontinuum. The MIR signal passes through a 2.4-$\mu$m long-pass filter (LPF), and its polarization is adjusted via a half-wavelength plate (HWP). The illumination power on the sample can be adjusted by a series of calibrated attenuators. In parallel, the pump pulse from the YDFL is sent through a chirped fiber Bragg grating (CFBG) to implement the time-stretch operation. The resultant chirped pump is combined with the MIR signal by a dichroic mirror (DM), and the mixed beams are focused into a chirped-poling lithium niobate (CPLN) crystal to perform the sum-frequency generation. After a short-pass filter (SPF), the upconverted beam is expanded by a 4f system to adapt the size of the holographic reflective grating. The spatially dispersed light is collected by an off-axis parabolic mirror (OAP), which is subsequently mapped onto a programmed DMD to perform the spectral encoding. Finally, the modulated beam is focused via   a plano-convex lens and a cylindrical lens (CL) onto a silicon photodiode. The measured intensity values and the associated projected patterns allow us to reconstruct the MIR spectrum. The involved electronic timing and data acquisition are handled by a controlling unit based on a field programmable gate array (FPGA).}
\label{fig1}
\end{figure*}

\begin{figure*}[t!]
\includegraphics[width=0.8\textwidth]{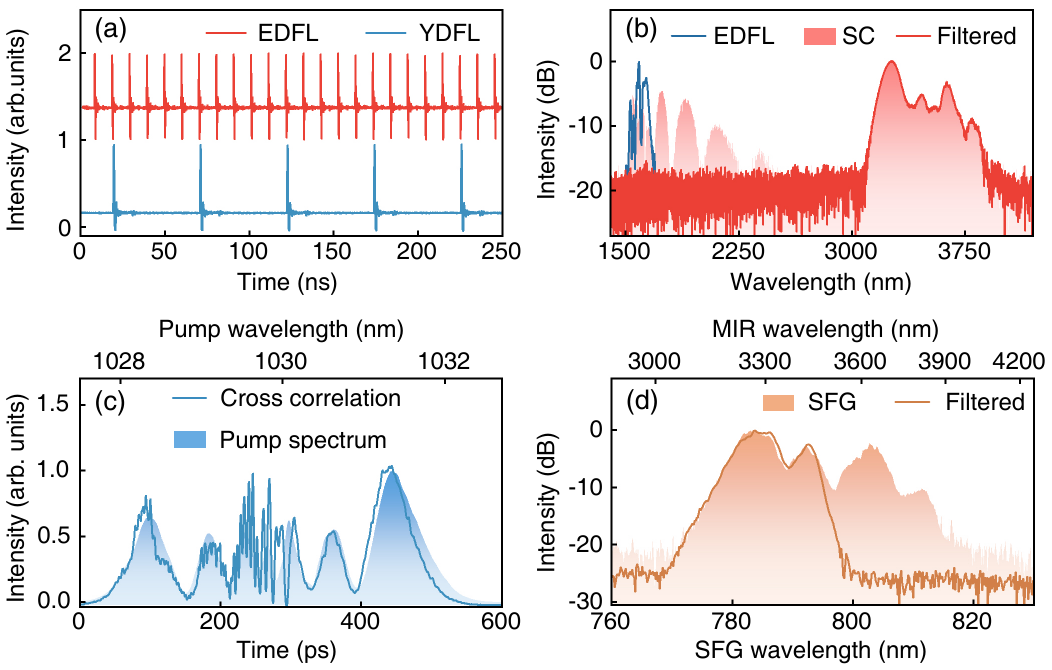}
\caption{Spectro-temporal characterization of involved optical pulses. (a) Recorded pulse trains from the synchronized EDFL and YDFL at repetition rates of 97 MHz and 19.4 MHz, respectively. (b) Measured optical spectra for the output pulse from the EDFL (solid blue line) before entering the Si$_3$N$_4$ waveguide, the generated MIR supercontinuum (shaded area), and the filtered light (solid red line) after a 2.4-$\mu$m long-pass filter. (c) Measured cross-correlation trace between the chirped pump pulse and the MIR signal, which can be regarded as the time-stretch profile of the pump. The temporal waveform is similar to the pump spectrum with a conversion factor about 120 ps/nm. (d) Measured optical spectra for the upconverted signal (shaded area) and the filtered light (solid line) after a short-pass filter at a cutting wavelength of 800 nm.}
\label{fig2}
\end{figure*}

The precise correspondence of the spatially dispersed spectral components to the DMD pixels is another prerequisite for approaching high-resolution spectrum reconstruction. Here, the upconverted beam is expanded via a 4f relay system to cover more grating area for making full use of the grating resolution. After being reflected from the grating, a sequence of spectral components are focussed onto the DMD surface, which spatially correspond to different electromechanical micromirrors \cite{Starling2016AO, Gattinger2022OE, Ebner2023SR}. The spectral resolution of the spectrometer is determined by the product of the focused beam size for a single wavelength component and the reciprocal linear dispersion (RLD) \cite{Xiang2011AS}. The RLD is defined as RLD = $\Delta \lambda/\Delta l$, where $\Delta \lambda $ is the difference between the wavelengths of two spectral lines and $\Delta l $ is the corresponding separated distance, which depends on the properties of the grating and the design of the optics.

The involved process of spectral reconstruction can be adapted from the single-pixel imaging \cite{Edgar2019NP}, which relies on the knowledge of the designated patterns and measured intensities. Specifically, we assume that the spectral intensity to be measured is given by $x(\lambda) \in \mathbb{A}^{N \ast 1}$, where $\lambda$ is the wavelength of the upconverted light, and $N$ is the number of spectral channels. The measurement process can be described by
\begin{equation}
y =  \Phi x,
\label{eq1}
\end{equation}
where $y \in \mathbb{A}^{N \ast 1}$ is the intensity recorded by the detector, and $ \Phi \in \mathbb{A}^{N \ast N}$ is the measurement matrix projected by the DMD. Usually, at least N measurements are required to recover the signal, since the measurement matrix must be inverted to achieve the spectrum reconstruction. Fortunately, most signals in nature are sparse when represented in an appropriate basis, which allows us to perform the sub-Nyquist sampling strategy. Such a compressive sensing facilitates the use of M (M$<$N) measurements to recover the signal without sacrificing too much information \cite{Edgar2019NP}. The involved operation is written as
\begin{equation}
y =  \Phi x = \Phi \Psi s = \Theta s,
\label{eq2}
\end{equation}
where $y \in \mathbb{A}^{M \ast 1}$, $ \Phi \in \mathbb{A}^{M \ast N}$, and $\Theta = \Phi \Psi \in \mathbb{A}^{M \ast N}$. $s \in \mathbb{A}^{N \ast 1}$ is the sparse vector that contains the projections of $x$ in the sparse sampling basis $\Psi \in \mathbb{A}^{N \ast N}$. When $\Theta$ satisfies the restricted isometry property (RIP), the sparse signal can in principle be reconstructed exactly \cite{Duarte2008IEEE}. Specifically, we use the Walsh-ordered Hadamard matrix as the measurement matrix $\Phi $ in the experiment, which ensures an incoherence to $\Psi$. The Walsh-ordered Hadamard transform is a generalized discrete Fourier transform, and hence the row vectors of matrix represent different frequencies in an increasing order \cite{Zhao2021Optica}. An efficient and accurate approach for signal recovery under the compressive measurement can resort to the algorithm based on total variation minimization \cite{Li2013COA}. More descriptions about the measurement matrices and reconstruction algorithms are given in Supplementary Note 2 and 3, respectively.

\begin{figure*}[t!]
\includegraphics[width=0.8\textwidth]{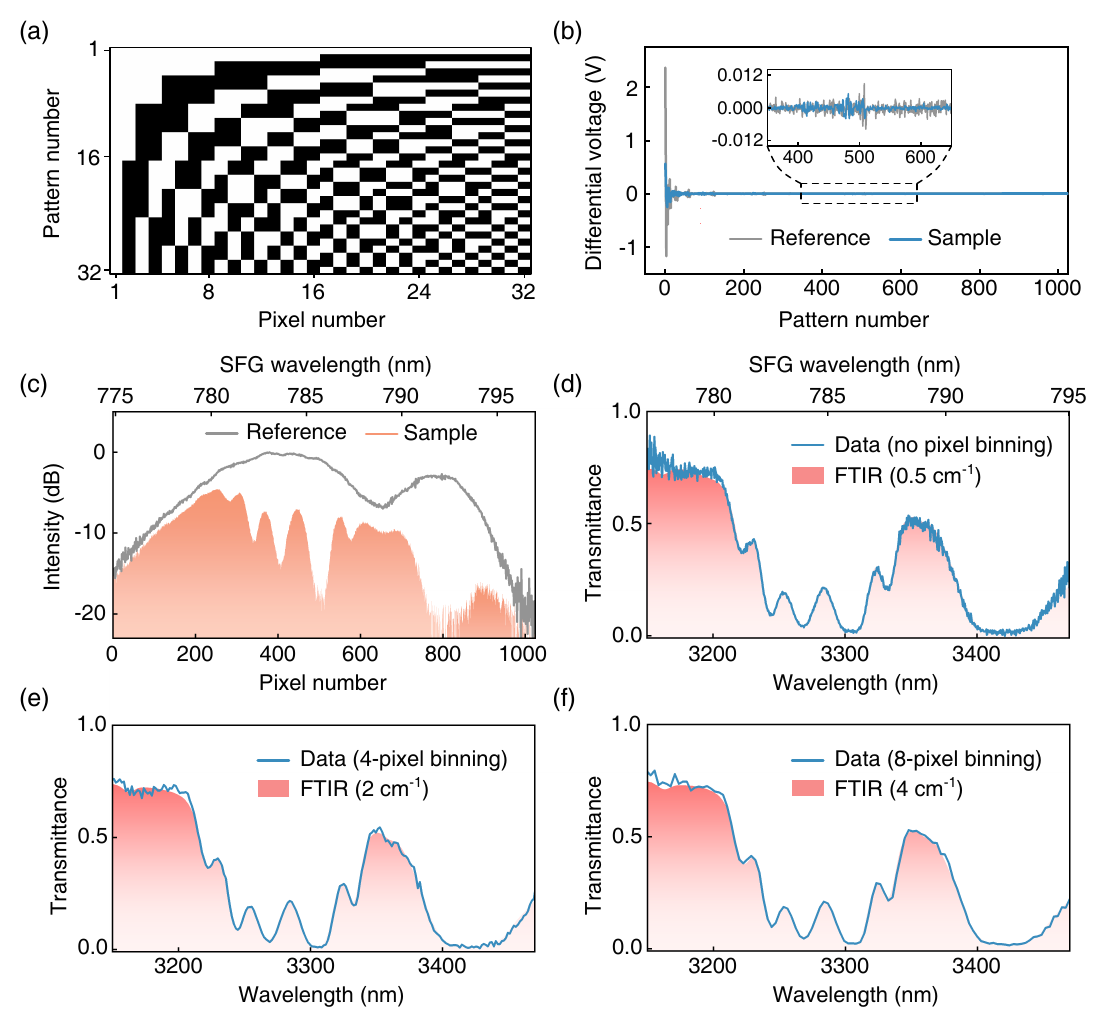}
\caption{High-resolution MIR spectral-coding spectroscopy. (a) A representative illustration for the 32-dimensional Walsh-ordered Hadamard matrix. (b) Measured differential voltages in the case of using two complementary patterns based on the 1024-dimensional Walsh-ordered Hadamard matrix, with and without the presence of the sample. Inset shows the detail of the small voltage variations for different patterns. (c) Reconstructed upconversion spectrum (shaded area) at the presence of a polystyrene film with a thickness of 50 $\mu$m. The reference without the sample is presented by the solid line. (d-f) Inferred MIR absorption spectra from the measured pixel-wavelength relationship. These spectra are acquired by operating the DMD at the conditions of no binning (d), 4-pixel binning (e), and 8-pixel binning (f), which agree well with the measured FTIR traces at spectral resolutions of 0.5 cm$^{-1}$, 2 cm$^{-1}$, and 4 cm$^{-1}$, respectively. }
\label{fig3}
\end{figure*}

\section{Experimental setup}
Figure \ref{fig1} illustrates the experimental setup of the single-pixel MIR spectral-coding spectrometer. The involved light sources are prepared from a synchronized fiber laser system, which consists of an Er-doped fiber laser (EDFL, LangyanTech, ErFemto Elite) and an Yb-doped fiber laser (YDFL, LangyanTech, YbPico Elite) at repetition rates of 97 MHz and 19.4 MHz, as shown in Fig. \ref{fig2}(a). The optical synchronization is realized based on light injection from the master to slave lasers, which eliminates the stringent requirement of high-speed feedback electronics in conventional laser-cavity stabilization system \cite{Huang2021PR, Zheng2023LPR, Sun2024LPR}. Such a passive locking provides a simple yet robust way to obtain synchronized dual-color pulses at disparate wavelengths with a high timing precision. The EDFL at 1550 nm delivers ultrafast pulses with a temporal duration of 116 fs and an average power of 230 mW. The peak power is estimated to be about 20 kW, which favors the MIR supercontinuum generation in a nanophotonic Si$_3$N$_4$ waveguide (fabricated by LIGENTECH). The physical dimensions of the optical waveguide are specified to be 0.82 $\mu$m (height) $\times$ 2.75 $\mu$m (width) $\times$ 5 mm (length). The transverse size plays a key role in tailoring the overall dispersion of the waveguide, which is optimized to improve the flatness and efficiency of the MIR light through the soliton-induced dispersive wave generation \cite{Guo2018NP}. As depicted in Fig. \ref{fig2}(b), the generated MIR supercontinuum is concentrated in the region from 3100 to 3900 nm after a long-pass filter at a cutting wavelength of 2.4 $\mu$m. Hence, the prepared MIR illumination source covers a functional group region in spectroscopic analysis, including common single-bond vibrational modes such as C-H, O-H and N-H.

In parallel, the synchronized pump pulse from the YDFL at 1030 nm is coupled into a chirped fiber Bragg grating (CFBG, AFR, PSCG-1030-10.0-1) with a dispersion about 120 ps/nm. The time-stretch pump is then combined with the MIR pulse into a chirped-poling lithium niobate (CPLN, CTL Photonics, M1900-2400Chirp) to perform the sum-frequency generation (SFG). The CPLN crystal has a linearly increasing poling period from 19 to 24 $\mu$m along the length of 20 mm, and has a thickness and width of 1 mm and 3 mm, respectively. The involved adiabatic quasi-phase matching facilitates a broad and efficient nonlinear upconversion \cite{Zheng2023LPR, Sun2024LPR}. A free-space delay line is inserted in the pump path to adjust the temporal overlap between the dual-color pulses. Figure \ref{fig2}(c) presents the measured cross-correlation trace, which can be treated as the temporally stretched profile of the pump pulse. As expected, the pulse profile is matched to the optical spectrum of the pump. During the SFG process, the femtosecond MIR only interacts with a small fraction of the pump spectrum, which allows a high-fidelity spectral transfer during the nonlinear parametric mixing \cite{Jonusas2024OE}. In the experiment, the MIR pulse duration is inferred to about 110 fs \cite{Sun2024LPR}, and the pump pulse with a bandwidth of 3 nm is elongated to about 350 ps. Hence, the spectral resolution due to the SFG process is calculated to be about 0.01 cm$^{-1}$. Such a chirped pulse upconversion technique is the key to realizing the subsequent high-resolution spectroscopy. The corresponding upconversion spectrum is recorded as shown in Fig. \ref{fig2}(d). The spectrum is narrowed by a short-pass filter at a cutting wavelength at 0.8 $\mu$m, which facilitates the demonstration of high-resolution spectroscopy. Note that such a filtering operation is optional depending on the trade-off between the targeted bandwidth and resolution \cite{Xiang2011AS}.

Subsequently, the upconverted beam is expanded by a 4f imaging system that includes two lenses with focal lengths of 30 mm and 200 mm. The expanded beam diameter is about 20 mm to match the geometric size of a holographic reflective grating (Thorlabs, GH25-18V). The grating has a groove density of 1800 groove/mm, and has a maximum refraction efficiency about 80\% in the range of 500-1000 nm for the perpendicular polarization. The angularly separated spectral components from the grating are collected by an off-axis parabolic mirror (OAP, Thorlabs) with a focal length of 101.6 mm, which results in an array of focused spots on a DMD (Texas Instrument, DLP650LNIR). At the presence of a monochromatic illumination from a single-frequency laser diode, the diameter of the focused beam on the DMD is measured to about 10 $\mu$m, which is designed to match the pixel size of 10.8 $\mu$m of the DMD. The DMD is equipped with 1280$\times$800 pixels. The micromirrors can be tilted by on/off angles of $\pm$12$^\circ$ at a maximum switching rate of 10 kHz. The DMD is loaded with a sequence of predetermined patterns. Since the light is dispersed in the horizontal direction, the pixels in the same column are set to identical values, hence resulting in a programmed sequence of vertical strips. The modulated beam reflected from the DMD passes through a lens (f=50 mm) and a cylindrical lens (CL, f=50 mm) to focus the spectral-coding beam onto a silicon-based photodiode. Note that differential measurements of each mask and its inverse are performed to reduce the effect of the background noise and to suppress low-frequency oscillations for the illumination source, which help to improve the reconstruction quality \cite{Edgar2019NP, Wang2023NC}. The involved timing control and data acquisition are implemented by a high-precision digital module based on a field programmable gate array (FPGA). The operating configuration and timing sequence are detailed in Supplementary Note 1.

\begin{figure*}[t!]
\includegraphics[width=0.8\textwidth]{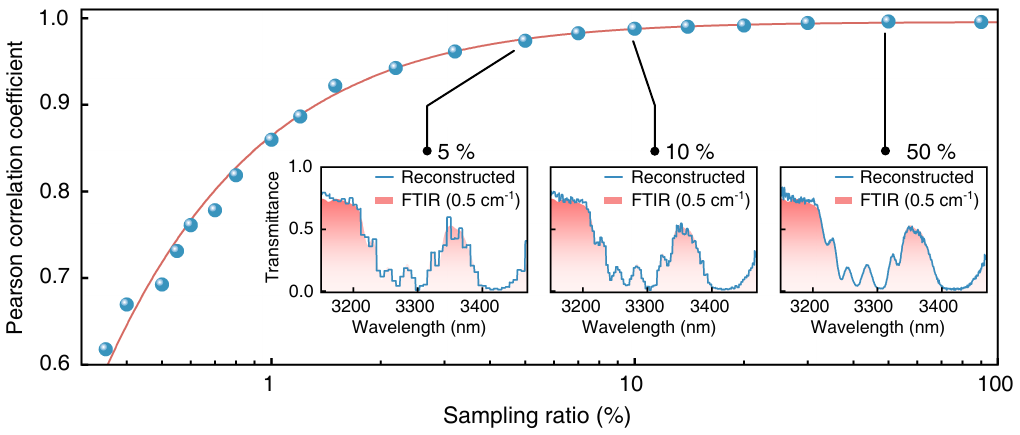}
\caption{Compressive MIR single-pixel spectroscopy. Reconstruction fidelity evaluated by the Pearson correlation coefficient (PCC) as a function of the sampling ratio. The solid line denotes an empirical fit with the exponential function. Inset presents the reconstructed MIR absorption spectra with sampling ratios of 5\%, 10\%, and 50\%.}
\label{fig4}
\end{figure*}

\begin{figure*}[t!]
\includegraphics[width=0.8\textwidth]{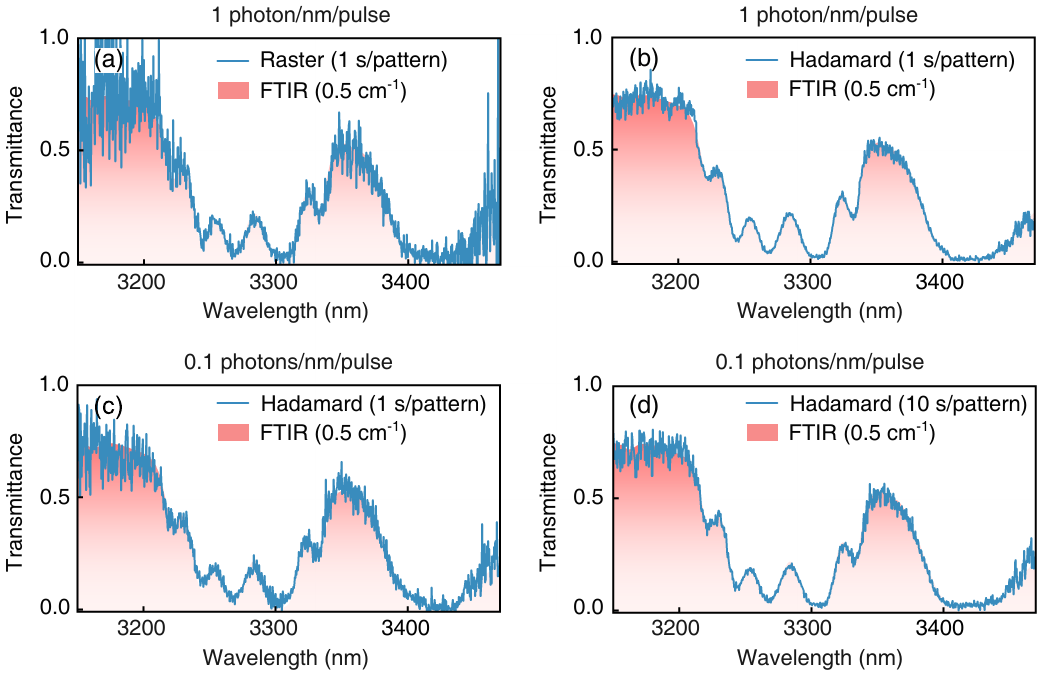}
\caption{Single-photon MIR computational spectroscopy. (a, b) Performance comparison for the reconstructed MIR absorption spectra in the cases of using raster scanning (a) and Hadamard sampling (b) under an illumination power of 1 photon/nm/pulse. The acquisition time is set to be 1 s for each pattern. (c, d) Reconstructed spectra in the case of a light flux of 0.1 photon/nm/pulse for acquisition time of 1 second (c) and 10 seconds (d) per pattern.}
\label{fig5}
\end{figure*}

\section{Results and discussion}
We start to calibrate the implemented MIR single-pixel upconversion spectrometer. To this end, an acousto-optic tunable filter (AOTF) is used at the output of the frequency upconverter. The AOTF is specified with a spectral bandwidth of 0.9 nm, and can be fast switched in 10 $\mu$s. For each filtered light, the DMD performs a raster scanning across all the pixels. The peak of the measured intensity trace can thus be used to identify the corresponding pixel for the given wavelength. Note that the short-pass filter for the SFG beam is removed for the calibration process. The two wavelengths for the first and last pixels can be identified, which lead to a measured RLD of 2 nm/mm. Consequently, the theoretical resolution of the system is calculated to be 0.4 cm$^{-1}$. The use of gratings with a higher line density could enhance the spectral resolution at the price of a smaller capture range, as commonly seen in grating-based spectrometers. Notably, the use of spatial modulators with more pixels would alleviate the involved performance trade-off. In the following experiment, the filter is placed back into the upconversion beam for characterizing the spectrometer performance, since the DMD can be almost occupied by the light with a spectral bandwidth about 22 nm.

Figure \ref{fig3}(a) presents the patterns for the 32-dimensional Walsh-ordered Hadamard matrix. All rows except the first one in the mask have half of the pixels in the ``on" state, thus showing the high-throughput advantage in comparison to the raster scanning approach \cite{Lim2021OE, Gattinger2021Sensors}. The actual matrix loaded to the DMD has a dimension of 1024. Here we we use an avalanche photodetector (APD, Thorlabs, APD440A) as the single-element detector, which features a high detection sensitivity with a minimum noise equivalent power (NEP) of 3.5 fW/$\sqrt{\text{Hz}}$. As discussed previously, the weighted signal is obtained by measuring the differential intensity for each pattern and its contrast inverse, which enables us to remove any offset in the spectrum due to background noise and suppress low-frequency oscillations for the illumination source \cite{Edgar2019NP}. The measured differential voltages are shown in Fig. \ref{fig3}(b) for various patterns with and without the presence of the sample of a polystyrene film with a thickness of 50 $\mu$m. The spike for the first pattern corresponds to the all-on state for the DMD. For the other patterns, the differential detection results in subtle changes of the signal around the mean value of zero. The reconstructed spectral profiles for the signal and reference are presented in Fig. \ref{fig3}(c). The corresponding MIR spectra can be inferred from the energy conservation law as $1/\lambda_\text{SFG} = 1/\lambda_\text{pump} + 1/\lambda_\text{MIR}$, where $\lambda_\text{pump, MIR, SFG}$ denote the wavelengths for the three interacting optical fields. Figure \ref{fig3}(d) shows the MIR absorption spectrum for the sample, which agrees well with the benchmark measured by a commercial FTIR spectrometer (Spotlight 400, PerkinElmer) at a spectral resolution of 0.5 cm$^{-1}$. Similarly, the DMD pixels can be combined to form a superpixel. The resulting Hadamard matrix with a smaller dimension reduces the number of intensity measurements for the spectral reconstruction, at the price of a degraded resolution. Figures \ref{fig3}(e) and (f) present the reconstructed spectra in the case of 4-pixel and 8-pixel binning, which are compared with the FTIR references at resolutions of 2 and 4 cm$^{-1}$, respectively. The illumination time for each pattern is identical for the three binning cases. It can be seen that less effective pixels lead to a more smooth spectrum, since more photons are distributed to each individual superpixel. Such a binning operation is favorable for high-speed spectroscopy, that is related to investigate dynamic scenes with a tolerance to a relative coarse resolution. Notably, one unique feature for the digitally programmable masks is the possibility to selectively interrogate a particular section. Moreover, instead of a uniform sampling, the use of a flexibly warped trajectory allows to efficiently allocate more spectral samples to the area of interest  \cite{Phillips2017SA}. Such a foveated operation provides a potential solution to balance the high resolution and operation range in practice, which is inaccessible with conventional spectrometers.

\begin{figure}[b!]
\includegraphics[width=0.85\columnwidth]{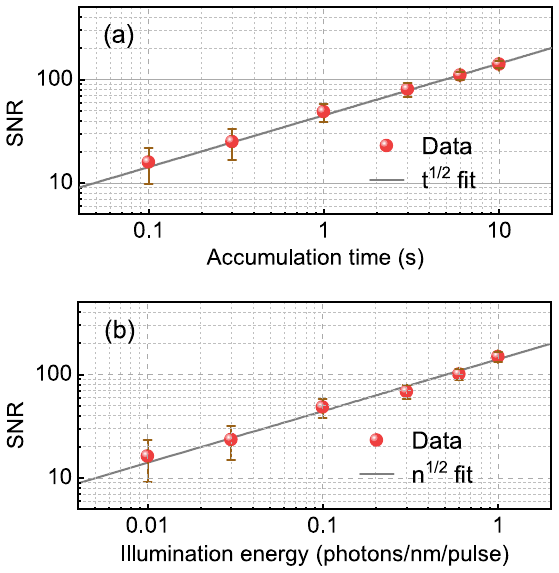}
\caption{Detection sensitivity for the MIR single-pixel upconversion spectrometer. (a) Signal-to-noise ratio (SNR) varies as the increase of accumulation time for each pattern under a constant illumination flux of 0.1 photons/nm/pulse. (b) SNR as a function of the incident pulse energy on the sample at a fixed accumulation time of one second for each pattern. Note that the fitted lines are given by the square root of the variables. The error bars denote standard deviations for five repetitive experiment measurements.}
\label{fig6}
\end{figure}

Next, we turn to demonstrate a compressive MIR single-pixel spectroscopy, which provides an alternative way for recovering high-resolution spectra in an efficient sampling manner. In the above approach, the number of measurements is required to be identical to the pixel number of the reconstructed spectrum. In contrast, a more sophisticated strategy based on the compressive sensing algorithm allows us to stably reconstruct the targeted spectrum from fewer samples or measurements while preserving its quality \cite{Gattinger2021Sensors, Starling2016AO, Gamez2016JAAS}. The underlying mechanism lies in the prior knowledge of the signal sparsity and incoherence, which enables the formulation of an optimization problem to reconstruct the signal by using a reduced number of sampling points below the Nyquist-Shannon criteria \cite{Duarte2008IEEE}. Experimentally, we randomly select M rows from the Walsh-ordered Hadamard matrix. By correlating the projected patterns and corresponding signal recorded on the detector, a N-element spectrum can be reconstructed via the total variation augmented Lagrangian (TVAL3) reconstruction algorithm \cite{Starling2016AO, Li2013COA}, as detailed in Supplementary Note 3. The quantity describing the level of compression in data acquisition is termed as sampling ratio, which is defined as the ratio between the number of measurements M taken in the experiment over N, \textit{i.e.}, measuring all the rows of the Hadamard matrix. Meanwhile, the Pearson correlation coefficient (PCC) is adopted to quantitatively measure the similarity between the recovered spectrum and the ground truth given by the FTIR trace. Figure \ref{fig4} presents the dependence of PCC on the sampling ratio, which indicates a high-fidelity reconstruction above 90\% with a small number of measurements down to 2\%. For sampling ratios over 5\%, the PCCs are kept above 97\%, which are manifested by the reconstructed traces in the inset. As discussed in Supplementary Note 4, there exist other figure of merits for evaluating the reconstruction quality, such as the mean square error (MSE), and the structural similarity index (SSIM). All these indicators verify the effectiveness of the reconstruction algorithm. The implemented compressive sensing operation favors a significant reduction of acquisition time, and thus facilitates a fast spectroscopic analysis. For instance, under the sampling ratio of 5\%, the number of  required patterns is $2\times1024\times5\% \approx 102$, which results in a total measurement time about 10 ms at a frame rate of 10 kHz for the DMD. Notably, emerging advances in single-pixel imaging have demonstrated pattern projection rates over MHz based on spinning cyclic masks \cite{Hahamovich2021NC} or swept aggregate patterns \cite{Kilcullen2022NC}, which would boost the frame rate of the MIR compressive spectrometer up to tens of kHz.

\begin{table*}[t]
\renewcommand\arraystretch{2}
\setlength{\tabcolsep}{8pt}
\caption{Performance comparison of broadband MIR single-photon spectrometers.}
\label{tab1}
\begin{tabular*}{0.8\linewidth}{@{}cccccc@{}}
\hline
 Ref. & Scheme & Range [$\mu$m] & Resolution [cm$^{-1}$] & Sensitivity &  Detector \\ 
 \hline
 This & SFG & 3.1-3.9 & 0.5 & 0.01 photons/nm/pulse & single pixel \\
\cite{Sun2024LPR} & SFG & 2.4-4.2 & 0.5 & 0.14 photons/nm/pulse & single pixel \\ 
\cite{Zheng2023LPR} & SFG & 2.4-4.2 & 5 & 0.2 photons/nm/pulse & array \\
\cite{Cai2022PR} & SFG & 3.1-3.8 & 10.5 & 0.09 photons/pulse & \quad single pixel$^{[a]}$ \\ 
\cite{Kalashnikov2016NP} & \quad SPDC$^{[b]}$ & 4.1-4.5 & $\approx$27 & \quad /$^{[c]}$ &  array \\  
\cite{Lindner2021OE} & SPDC & 3.1-4.0 & 0.56 & 11.9 photons/nm/pulse & \quad single pixel$^{[d]}$ \\
\cite{Tashima2024Optica} & SPDC & 2.0-5.0 & 4 & \quad /$^{[c]}$ & \quad single pixel$^{[d]}$ \\  
\hline
\end{tabular*}
\begin{tablenotes}
\item[] $^{[a]}$ A monochromator is used to conduct the spectral analysis based on a grating-scanning operation.
\item[] $^{[b]}$ This approach relies on quantum interferometry based on spontaneous parametric down conversion (SPDC).
\item[] $^{[c]}$ The MIR illumination power is not specified, but its sensitivity has reached the single-photon level.
\item[] $^{[d]}$ A mechanical delay scanning is required for the FTIR-type spectrometer.
\end{tablenotes}
\end{table*}

In the following, we proceed to characterizing the MIR computational spectroscopy at the single-photon level. The single-photon spectroscopy is manifested by a weak illumination spectral intensity at the single-photon level. To this end, the detector is replaced by a single-photon counting module (SPCM) based on a Geiger-mode avalanche photodiode (Laser Components, SAP500T8). The diameter of the optical sensor is 0.5 mm, and the detection efficiency is specified to be 54\% at 800 nm. To emulate the photon-starved scenario, the MIR illumination is attenuated by a series of calibrated neutral density filers. Different from the analogue detector, the output signal from the single-photon detector is a sequence of pulses that are responses to the incident photons. Hence, a frequency counter is used to record the event count within a customized integration time. At an illumination flux of 1 photon/nm/pulse, the reconstructed spectra in the case of using the raster scanning and Hadamard masks are given in Figs. \ref{fig5}(a) and (b), respectively. As expected, the raster scanning is inefficient to use the available light, which results in a noisy spectrum due to small signals emanating from a single aperture. This issue is more prominent at the low-light-level illumination or with a smaller aperture size for increased number of pixels \cite{Edgar2019NP}. In contrast, the spectral-coding modality favors the multiplex advantage to minimize the effect of detector noise by using more light in each measurement, which in principle can improve the SNR by the value $\sqrt{N/2}$, with $N$ being the number of DMD super-pixels \cite{Gattinger2021Sensors}. The factor of 2 is taken into account due to the two-fold measurements in the differential detection scheme. Here, the SNR is defined by the absorption peak at 3268 nm over the standard deviation of the background. The SNRs for the spectra in Figs. \ref{fig5}(a) and (b) are measured to be 6.5 and 149.2, which is consistent to the expected enhancement. As shown in Fig. \ref{fig5}(c), further reduction of the light flux to 0.1 photons/nm/pulse is still permitted for the Hadamard encoding to obtain a faithful spectrum with a SNR of 48.7. The SNR can be improved to 140.2 by resorting to a longer accumulation time up to 10 s per pattern as given in Fig. \ref{fig5}(d), which is close to the one obtained in Fig. \ref{fig5}(b) due to identical photons collected for each pattern.

Finally, we focus on the investigation of the detection sensitivity for the MIR single-pixel spectrometer. Figure \ref{fig6}(a) presents the measured SNR as a function of the accumulation time for each projected pattern. The incident photon flux is set to be 0.1 photons/nm/pulse, which corresponds to an illumination power about 2.3 aJ/pulse on the sample. In the case of a constant pattern displaying time of 1 s, the SNR increases with a higher illumination power, as shown in Fig. \ref{fig6}(b). A modest SNR about 16 can still be attainable at an illumination flux as low as 0.01 photons/nm/pulse. Specifically, the obtained SNR here is obtained for a total acquisition time of 1024 s with a sampling ratio of 50\% for the compressive measurement. Under comparable SNRs and acquisition time, the illumination flux is found to be about 0.1 photons/nm/pulse in the previous demonstration \cite{Sun2024LPR}. We note that the presented single-pixel spectrometer here indicates shot-noise-limited SNRs with a square root dependence on the illumination power, which would offset the multiplex advantage comparing to a scanning monochromator under the same dominant noise \cite{Griffiths2007Book}. However, the low-light-level illumination in the single-photon spectroscopy renders the scanning scheme operating in the detector-noise-limited regime. It is the detector noise suppression and the dominant noise regime shift that enable the multiplex advantage even under shot-noise-limited SNRs in our system. In the experiment, the conversion efficiency is measured to be about 0.4\% at a maximum pump power of 0.9 W. The overall detection efficiency is inferred to be about 0.03\% by including other loss sources from the transmission, filtering, and detection. Consequently, the detected light flux onto the optical detector is calculated to be about 2.4$\times$10$^{-3}$ photons/pulse. Such a low detected energy benefits from the low-noise nonlinear upconversion and high-sensitivity silicon detector. 

As summarized in Table \ref{tab1}, the achieved detection sensitivity represents at least a ten-fold improvement over reported values for previous demonstrations of broadband MIR single-photon spectrometers. Our implemented spectrometer eliminates the need of the single-photon detector array in the grating-based scheme \cite{Zheng2023LPR}, and alternatively offers an effective way to approach a high spectral resolution due to the readily available spatial modulator with massive pixels. Moreover, the encoding flexibility inherent to the digital modulator can facilitate programmable measurements within selective regions of interest. In combination with the compressive operation, the single-pixel computational spectrometer is beneficial to perform fast and sensitive spectroscopic analyses, which is particularly advantageous in photon-starved scenarios where a long accumulation time is needed for collecting sufficient photons to obtain high-contrast spectral traces. In comparison to the time-stretch scheme \cite{Sun2024LPR}, our system avoids the lossy channel of the long-haul fibers, thus resulting in more detected photons on the detector. It is the resultant higher system efficiency that leads to the sensitivity improvement. Therefore, our presented work not only establishes an effective path to realize ultra-sensitive MIR single-pixel spectroscopy, but also represents a landmark in high-resolution MIR single-photon spectrometers.

\section{Conclusion}
In conclusion, we have implemented for the first time a MIR single-pixel compressive spectrometer at the single-photon level, which combines the merits from frequency upconversion detection and single-pixel spectroscopy. In contrast to previous MIR upconversion spectrometers \cite{Zheng2023LPR, Sun2024LPR, Cai2022PR, Kalashnikov2016NP}, the presented scheme favors combined features of wavelength-multiplexing detection and high optical throughput while maintaining the single-pixel simplicity. In this architecture, classical moving parts for scanning, such as rotational gratings or translational stages, become unnecessary, and highly sensitive single-element detectors can be used as opposed to linear arrays. Consequently, the presented infrared spectrometer overcomes the stringent requirement for high-definition array detectors in precise spectroscopic measurements, and allows for sub-wavenumber resolution over a wide spectral coverage based on a single-photon detector. 

Furthermore, the proposed single-pixel upconversion spectrometer can also be readily extended to longer infrared \cite{Rodrigo2021LPR} or terahertz wavelengths \cite{Fandio2024OL} by resorting to nonlinear crystals based on silver gallium sulfide (AgGaS$_2$) and gallium phosphide (GaP), where high-performance detectors and optical modulators are currently hard to access in these highly-demanded fingerprint spectral regions. Additionally, in combination with the compressive upconversion imaging technique \cite{Wang2023NC}, an infrared single-pixel hyperspectral imager can be envisaged to acquire multidimensional information in a fast and sensitive fashion \cite{Wang2023NC, Spencer2016NC, Ebner2023SR}, which would provide a powerful tool for applications in materials analysis and chemical sensing at the photon-starved scenarios.

\section*{Acknowledgements}
This work was supported by National Key Research and Development Program (2021YFB2801100), National Natural Science Foundation of China (62175064, 62235019, 62035005, 12374313, 62175152); Shanghai Pilot Program for Basic Research (TQ20220104); Natural Science Foundation of Chongqing (CSTB2023NSCQ-JQX0011, CSTB2022NSCQ-MSX0451, CSTB2022TIAD-DEX0036); Shanghai Municipal Science and Technology Major Project (2019SHZDZX01); Fundamental Research Funds for the Central Universities.

\section*{Conflict of Interest}
The authors declare no conflict of interests.

\section*{Supporting Information}
Supporting Information is accompanied to present more details on the data acquisition and processing. 

\section*{Data Availability Statement}
The data that support the findings of this study are available from the corresponding author upon reasonable request.

\section*{Keywords}
mid-infrared spectroscopy, single-pixel spectroscopy, compressive sensing spectroscopy, single-photon spectrometer, frequency upconversion detection

\end{document}